# A Survey on Smart Metering Systems using Blockchain for E-Mobility

Juan C. Olivares-Rojas, Enrique Reyes-Archundia, José A. Gutièrrez-Gnecchi, Ismael Molina-Moreno
*División de Estudios de Posgrado e Investigación*
*Tecnológico Nacional de México/I.T. de Morelia*
Morelia, México
{juan.or, enrique.rm, jose.gg3, ismael.mm}@morelia.tecnm.mx

*Abstract*— Electricity is an essential comfort to support our daily activities. With the competitive increase and energy costs by the industry, new values and opportunities for delivering electricity to customers are produced. One of these new opportunities is electric vehicles. With the arrival of electric vehicles, various challenges and opportunities are being presented in the electric power system worldwide. For example, under the traditional electric power billing scheme, electric power has to be consumed where it is needed so that end-users could not charge their electric vehicles at different points (e.g. a relative's house) if this the correct user is not billed (this due to the high consumption of electrical energy that makes it expensive). To achieve electric mobility, they must solve new challenges, such as the smart metering of energy consumption and the cybersecurity of these measurements. The present work shows a study of the different smart metering technologies that use blockchain and other security mechanisms to achieve e-mobility.

## I. INTRODUCTION

The efficient use of electric energy has become a global necessity and Mexico is no exception. One of the approaches to achieve energy efficiency is to minimize losses in the different types of processes in the generation, transmission, distribution, and marketing chain. Within the commercialization, various losses of the non-technical type can be presented, such as fraud and energy theft, as well as measurement and billing errors [1].

The development of electric vehicles (EV) has brought several opportunities and challenges. Undoubtedly, the main benefits lie in apparent less pollution by stopping using fossil fuels such as gas and gasoline, but it also represents some problems such as where to charge electric vehicles. If the charging is done at the users' premises, e-mobility is lost, having to return to their home to charge. For this reason, public charging points similar to today's gas stations called electric charger station are needed. One of the possibilities could be to charge electric vehicles in private facilities such as the neighbors' house. In the past, this would not be a problem if the electric charges were small, as in the case of charging a cell phone, but now, these charges are greater and expensive for the friendly neighbor who lent us his facilities to charge our devices [2].

In general, energy consumption in developed countries is slowing and decoupling demand. Electricity consumption in the United States and the European Union is 40% worldwide, while in other countries especially emerging economies continue to grow [3]. Electric transport and e-mobility will be changing this growth paradigm. It is not only about generating more electricity but generating it more efficient and less polluting. Electricity generation can be seen as a complex multi-energy system: gas, coal, oil, nuclear, etc., it depends on various sources and costs.

Transportation is responsible for 28% of total energy consumption and 23% of $CO_2$ emissions. By 2060 the energy consumption due to transport will double. On the other hand, the industry is responsible for the consumption of 38% of energy and emits 24% of the total pollutant emissions. In 2015, China surpassed the United States in the use of EVs. According to the EVI 2020 initiative, 20 million electric vehicles around the world are expected in 2020 [4]. Global consumption of 2% is estimated due to the use of a data center, so not all digitalization will result in a decrease in electricity consumption. The Internet of Things (IoT) is bringing a considerable amount of new small electronic devices connected to the Internet and consuming electricity on a growing scale.

In order to achieve adequate mobility of EV chargers and other large loads of possible appliances/devices, it is necessary to be able to measure this consumption and invoice it, so that smart metering systems can be a good option.

Below is a review of the state of the art of the various existing mechanisms for measuring the consumption of an electric vehicle and its proper billing through the use of intelligent measurement systems.

## II. LITERATURE REVIEW

The smart grid has evolved in recent years, but it owes its great boom to the invention of smart electric meters. A smart electric meter allows the measurement of consumption and/or production of electrical energy and report it through telecommunications networks to the utilities, facilitating the process of reading the measurements and their rapid billing [1]. In addition to these benefits, there are also others such as cuts and reconnections automatically, periodic reports of energy consumption, interconnection with energy management systems and demand response [2].

In order to achieve all this, it is required that in addition to the smart meters of other components and technological infrastructure, the most widespread being AMI [3]. AMI allows data to be collected at various levels through data concentrators in addition to having a data center in the electricity company to be able to work with the large volume of information generated by smart meters [4].

Several authors have focused on providing new functionality to meters, particularly focused on cybersecurity [5-8], and data analytics for diverse applications [9-13], to name a few. Recently it has been of interest to look for methods that contribute to e-mobility, using various mechanisms.

Several authors have been responsible for conducting studies related to the integration of electric vehicles with the Smart Grid (SG). For example, [14] shows a study of how the various renewable energy technologies are used to create the concept of Vehicle to Grid (V2G), in which, the energy produced in excess can be stored for use in periods of high demand.

The most critical component in the infrastructure of EVs is the charging infrastructure and many of the works focus on this tenor. Work has also been done to extend the concept of EVs using hybrid energy cogeneration technologies if there is no electricity available [14].

Particularly in [15] the integration of the EV infrastructure with AMI is studied. This integration is not new and has been perfected for many years. This work mentions the InovGrid project developed in Portugal, where smart meters also help measure the consumption of EVs in their microgeneration.

Other works such as [16] have focused on the use of blockchains to improve loading station selection processes. To achieve energy consumption/production measurements, the system must be decentralized. Smart contracts have been used in a way that guarantees the confidentiality of energy transactions. It is important to record the state of charge in the percentage of the battery. The concept of digital currencies as cryptocurrencies has been used for energy transactions. One of the systems that have used blockchain is Plugshare, as an electronic payment system.

A review of the state of the art of different blockchain applications in SG is shown in [17]. Mainly it has been used in a P2P system to prevent fraud. The poor coordination of the use of EV charging stations causes big problems. Many works have focused on reducing power fluctuations and freight costs.

[18] shows a blockchain-based firmware scheme for autonomous vehicles. [19] shows a blockchain-based sustainable electricity energy ecosystem for prosumers where V2H (Vehicles to Home) and V2B (Vehicles to Building) are interrelated.

[20] shows a study of a probabilistic approach combining smart meters and charging data of EVs to investigate its impact on the distribution network. The Project developed is called SwitchEV, where the EV is the primary vehicle.

In [21] the application of blockchain technology in the fields of Internet Energy and its need for use is introduced. [22] shows the use of computation in the fog through the use of a blockchain-based vehicular assistant for carpooling.

[23] a blockchain-based energy marketing platform for smart homes and microgrids is presented. This work uses a home miner in charge of monitoring energy use in real-time.

[24] shows the use of SGs based on blockchain for the use of local sustainable energy markets. Local energy markets are where consumers and prosumers exchange energy in their community.

Other related work to electricity markets is presented in [25]. This paper presents the potential impact of using blockchain-based solutions for energy markets. There are various billing and payment schemes for EV charging. Sharing information between charging stations is extremely complex. The optimization of the electricity markets translates into potential cost savings. A small saving in EV translates into a considerable amount due to its large increase in use.

On the other hand, [26] shows a feasibility analysis of a decentralized energy market based on blockchain. For this, they use the Ethereum platform and the real-time simulator GridLAB-D.

In [27] an alternative scheme to obtain energy quality information is presented through the study of load signatures and their applications. Each electrical device can be characterized differently through its consumption patterns.

[28] shows the formulation of load signatures through monitoring based on non-intrusive current measurements. This paper reviews the 1st, 3rd and 5th harmonics using a methodology for data processing.

The signature knowledge of electric charges is the principal or basis of practical charging monitoring technologies, which involve the identification of household appliances and the determination of their operating status [29]. This knowledge can serve public utility companies, customers, appliance manufacturers, and other stakeholders.

In [30] the automatic identification of household appliances charging signatures using statistical groupings is presented; while in [31] a small review of the state of the art of technologies, business models and adaptation of market design for smart electricity distribution is presented.

[32] shows a study with the political implications for electricity laws in the European Union through the change in the role of consumers of electric energy through the use of blockchain.

[33] a study of consensus mechanisms in blockchain-based on credits is presented. This algorithm is a variation of PBFT (Practical Byzantine Fault Tolerance); while in [34] the DeepCoin cryptocurrency is presented, which is based on a deep learning scheme to achieve consensus in their transactions.

[35] shows a study on the main blockchain mechanisms implemented in SG. Among the main projects are toblockchain from the Netherlands, Spherity from Germany, Solara from Australia, Share & Charge Foundation from Germany.

[36] shows a study for the electrical transition using blockchain technologies; while [37] and [38] show a study on the various digital energy solutions through blockchain for a decentralized and decarbonized sector.

[39] shows a forecast study in electricity markets using blockchain technologies. Local electricity markets will be a trend with the use of electric microgrids.

[40] shows a study of the energy transition change and the use of blockchain in smart cities. The blockchain can facilitate e-mobility as a service through the realization of a simple and cheap power charger. It can also be "uberized" by sharing the use of electric vehicles and their recharging among different users. For example, Sunchain is one of the blockchain systems created in this tenor.

[41] shows a study of the various obstacles that must be overcome for the proper implementation of e-mobility in Spain. A possible solution to the recharging scheme is to place chargers on medium and high voltage pole. It is proposed to charge electric vehicles at night, but a broader study of the peaks and ridges of electricity consumption has to be done.

[42] shows an executive study of the main blockchain applications in the electricity sector; while in [43], a study on different cases of blockchain use in the electricity sector is shown. A case study is the infrastructure for peer-to-peer energy exchange. Another popular case is the exchange of energy in electric vehicles. For example, unused electric vehicles can be discharged to power other vehicles that do. Another popular case is the use of techniques to maintain security and privacy.

[44] shows the use of blockchain in various contexts of the SG, particularly in forensic matters. In turn, [45] a study of electric mobility for Latin America is presented. In many Latin American countries, there are fossil fuel subsidies that make the generation of electricity economically profitable through these non-renewable means.

In [46] a whole study on the blockchain revolution in the energy sector is shown, while in [47], the fifth archetype in the power grid revolution is presented: the use of blockchain in the electricity markets. The 4 archetypes of the electricity grid have been: 1. The centralized utility model. 2. The disaggregated retail model. 3. The platform model. 4. The balanced operator model.

The passive consumer has not only become a producer of electricity but has also adapted to demand response programs depending on the generation capacity of the electricity grid [32]. In some countries of the European Union, users are allowed to feed energy produced to the network only if it is free. RES management involves the use of flexible technologies such as storage, electric vehicles, smart home technologies; since the power grid is not designed in two ways.

Blockchain technologies allow small producers to enter the electricity market. It is intended to increase the number of electricity generation by renewables by significantly reducing pollution from the use of fossil resources. The market will be more dynamic and more participatory, there will be greater segmentation of small electricity markets. New financing mechanisms are expected. Traditional contracts are slow and costly. Smart contracts will revolutionize this area. Blockchain works are more focused on finance than the energy part [36].

The blockchain can be used in all phases of the electric power process: generation, transmission, distribution, and marketing. For many authors, the blockchain disruption resembles that of the Internet 30 years ago [37]. For the proper functioning of an energy system, data is needed that allows its operation and improvement. The best mechanisms to eliminate the variation of renewable energies are: use storage systems, make forecasts of energy systems, design of electricity markets, demand management, regional integration. The objective of any technology in SG is to make the SG more flexible, in order to maintain the reliability and security of the network, as well as avoid the volatility and variability of renewable energies. The vast majority of blockchain jobs are made in Europe, the United States, and Australia. However, Latin America is a region with abundant renewable energy. Various DLT technologies have been used for payment systems particularly in charging stations. Blockchain is a core element for the energy systems of the future. It has been proven that a complete decentralization of blockchain is not functional for electric companies, the best example is Bitcoin. Traceability is extremely important for the certification of clean energy.

46% of Blockchain initiatives for energy are in Europe [43]. The three countries with the most developments are the United States, Germany, and the Netherlands. The most common use of blockchain is a peer-to-peer exchange of energy. About 50% of projects use Ethereum. The energy consumption of blockchain is estimated at 0.2% of the world's total.

A load signature is the unique behavior of individual devices that can be disaggregated from the components of the load signal [27]. Signatures can be identified on two levels: micro and macro. The micro-level has to do with sampling less than a second. Various techniques have been used to decompose signals such as FFT (Fast Fourier Transform), high-order harmonics in steady-state signals, wavelet transform and geometric characteristics of waveforms.

The times and places of the load of the measurements should be considered, since in many places the rates are not flat, in addition to having incentives to make the cost of electricity for the recharging of EVs cheaper. There are different approaches to understand the characteristics of electric charge, among the main ones, are [29]:

1. Methods to measure and represent load characteristics.
2. The Development of signal processing techniques and estimation algorithms for signal filtering, signal disaggregation, and load recognition.

In addition, there are two types of load models [15]:

1. Without control: Silly load and MPT (Multi-Process Rate).
2. With control: Smart charging and V2G.

With the disaggregation of load signals at the smart meter level, new applications can be obtained. Load unbundling is breaking the composition of a load signal into an identifiable set of devices [30]. Once the identification is done, it is possible to follow a specific appliance and record its usage pattern in detail. In general, there are two approaches to solving the problem of disaggregation: optimization or pattern recognition. Creating a database of electronic signatures requires many heuristic rules or human interpretations and it is not feasible to build the database for each appliance in millions of houses. A charging signature is defined as an electrical behavior of an individual equipment/appliance when it is in operation. A composite load (CL) is a combination of multiple load signatures at a specific point. Active Power (P) and Reactive Power (Q) are two important parameters for finding load signatures. The STC is the transient current change and can also help identify individual appliances.

In addition, EV charging systems must consider various aspects such as periodic reports, data storage, connectivity, privacy and security, data encryption, logging, clock, firmware update. The selection of the type of contract. Failure communication. Electric roaming is extremely important.

The review of the aforementioned literature has not found the development of a smart metering system that integrates the use of electric charge identification in smart meters using a blockchain to allow e-mobility and encourage users to do it with a social responsibility approach to the environment.

## III. COMPARATIVE ASSESMENT

In Table I are the main works with their most relevant characteristics for their application in smart metering systems in order to achieve easy e-mobility.

Among the main desirable features in a smart meter for mobility management is blockchain management, the possibility of integrating into local energy markets as well as load signature detection methods. The latter has not been handled in current smart metering systems but from our point of view, they will be essential for the correct integration of the e-mobility and billing of any electrical device anywhere in the immediate future.

Table I. Smart Metering E-mobility Related Works.

| Work | E-mobility | SM | Blockchain | Load Signature | Market |
|------|------------|----|------------|----------------|--------|
| [16] | Y | N | Y | N | N |
| [18] | Y | N | Y | N | N |
| [19] | N | Y | Y | N | Y |
| [20] | Y | Y | N | N | N |
| [21] | N | Y | Y | N | Y |
| [22] | Y | N | Y | N | Y |
| [23] | N | Y | Y | N | Y |
| [24] | Y | Y | Y | N | Y |
| [28] | N | N | N | Y | N |
| [34] | N | N | Y | N | Y |

## IV. CONCLUSIONS

E-mobility will be a reality in the coming years due to the increasingly scarce fossil energy resources such as gas, coal, the oil used in transport systems. The SG must consider a scheme for measuring energy consumption and its respective billing. The present work showed a comparative review of the current state of the art of the works that are being developed in the field of smart metering with respect to achieving an efficient and environmentally friendly e-mobility.

ACKNOWLEDGMENT

The authors thank the Tecnológico Nacional de México for the support provided for the development of the Project under grants 7948.20-P and 8000.20-P.